# Acetaminophen Interactions with Phospholipid Vesicles Induced Changes in Morphology and Lipid Dynamics

Judith U. De Mel[1*], Sudipta Gupta[1*], Sydney Harmon[2], Laura Stingaciu[3], Eric W. Roth[4], Miriam Siebenbuerger[5] Markus Bleuel[6], Gerald J. Schneider[1,7*]

[1]Department of Chemistry, Louisiana State University, Baton Rouge, LA 70803, USA
[2]Department of Chemistry, Colorado School of Mines, Golden, CO 80401, USA
[3]Neutron Sciences Directorate, Oak Ridge National Laboratory (ORNL), POB 2008, 1 Bethel Valley Road, Oak Ridge, TN 37831, USA
[4]Department of Materials Science and Engineering and NUANCE Center, Northwestern University, 2220 Campus Drive, Evanston, Illinois 60208, USA
[5]Center of Advanced Microstructures and Devices, Louisiana State University, 6980 Jefferson Highway, Baton Rouge, LA 70806, USA
[6]NIST Center for Neutron Research, National Institute of Standards and Technology, Gaithersburg, MD 20899-8562, USA
[7]Department of Physics & Astronomy, Louisiana State University, Baton Rouge, LA 70803, USA

*Corresponding Authors: Judith U. De Mel, Sudipta Gupta, Gerald J. Schneider

## Abstract

Acetaminophen (APAP) or Paracetamol, despite its wide and common use for pain and fever symptoms, shows a variety of side effects, toxic effects, and overdose effects. The most common form of toxic effects of APAP is in the liver where phosphatidylcholine is the major component of the cell membrane with additional associated functionalities. Although this is the case, the effects of APAP on pure phospholipid membranes have been largely ignored. Here, we used DOPC, a commonly found phospholipid in mammalian cell membranes to synthesize large unilamellar vesicles to investigate how the incorporation of APAP changes pure lipid vesicle structure, morphology, and fluidity at different concentrations. We used a combination of dynamic light scattering (DLS), small-angle neutron and X-ray scattering (SANS, SAXS), cryo TEM for structural characterization, and neutron spin-echo (NSE) spectroscopy to investigate dynamics. We showed that the incorporation of Acetaminophen in the lipid bilayer significantly impacts the spherical phospholipid self-assembly in terms of its morphology as well as influences the lipid content in the bilayer, causing a decrease in bending rigidity. We discussed how the overall impact of APAP molecules on the pure lipid membrane may play a significant role in the drug's mechanisms of action. Our results showed the incorporation of APAP reduces membrane rigidity as well as changes the spherical unilamellar vesicles into much more irregularly shaped vesicles. Although bilayer structure did not show much change when observed by SAXS, NSE and cryo-TEM results showed the lipid dynamics change with the addition of APAP in the bilayer which causes the overall decreased membrane rigidity. A strong effect on the lipid tail motion was also observed.

## Introduction

Acetaminophen (APAP) and popular NSAIDs (Nonsteroidal anti-inflammatory drugs) such as Aspirin (ASA), and Ibuprofen (IBU) have been used for decreasing inflammation and pain relief for centuries. Commercially available from the 1950s, many Over-the-counter (OTC) anti-inflammatory and pain medications are extensively used worldwide without prescription control. Among these, Acetaminophen is the most commonly used which has a profound presence in antipyretic and analgesic usage that it is almost to a point of over-use.[1] The danger arises due to the toxic effects and side effects on mammals these drugs can have.[2-6] It has also been shown that despite the low dosages consumed, the drugs tend to accumulate and concentrate in different tissues where the therapeutic effects, side effects, and toxic effects are detected. Therefore, researchers from various disciplines have attempted to unravel their mechanisms of actions in humans and other animal and tissue models using a variety of approaches.

APAP or Paracetamol is considered the first-line choice for pain relief while drugs such as ASA and IBU are considered anti-inflammatory counterparts.[7] Despite the differences in application, one thing these drugs have in common is that their main mechanism of action is connected to a membrane-bound protein family called cyclooxygenase (COX) which regulates prostaglandin formation which then in return regulates inflammatory responses and pain.[8] Over the years, the focus of understanding underlying mechanisms of actions of APAP has been in connection to the COX (Cyclooxygenase) enzyme centers and other related proteins. Despite the long history of use in modern medicine, mechanisms of action of APAP are complicated and not completely understood.[9-12] Although this is the case, APAP is currently known as a COX inhibitor by competitive inhibition to the active site which binds arachidonic acid,[9] as opposed to the action of other drugs such as IBU which is related to non-specific inhibition of COX[13], or ASA which has shown activity related to chemopreventive effects and platelet aggregation in addition to being a COX inhibitor by covalently modifying COX active site.[14] Therefore, despite the common relationship with the COX enzymes, these drugs have diverse interaction pathways which call for an independent investigation of the impacts of the drugs to understand their unique effects. One aspect that requires continuous attention is the drug-lipid membrane interactions. Studies have shown different impacts of small drug molecules on lipid membranes such as induced fusion[15], membrane permeability[16, 17], and changes in membrane rigidity.[18-20] Particularly, drug-induced membrane rigidity changes are important to explore due to the relationship of membrane rigidity with multiple functions such as the metastatic potential of cancer cells[21-23], apoptosis[24], erythrocyte morphology[25], etc.

Acetaminophen is a small drug molecule ($C_8H_9NO_2$, 151.1626 g mol$^{-1}$) that consists of an aromatic core with an acetanilide functional group and a hydroxyl functional group in para positioning to each other. The pKa value of 9.38 renders the molecule to be charge-neutral in physiological pH as shown in Figure 1.[26] Molecular dynamics simulations predict APAP molecules are located close to the carbonyl group region of the phospholipids, in an intermediate location between the hydrophobic and hydrophilic parts of the phospholipid.[27] Understanding the unique effects of APAP on mammalian cells particularly, molecular-level details of the drug's influences on physicochemical properties of the cell membrane integrity and fluidity is critical for future therapeutic advancements (e.g. decreasing side-effects due to toxicity). APAP overdose effects have also been researched extensively.[28] The most common form of APAP toxicity occurs in the liver where phosphatidylcholine (PC) is the primary component of the cell membrane in addition to its other functions such as being a precursor of signaling molecules as well as being a key element in lipoprotein and bile.[29] Many studies have shown a connection between phospholipids with the APAP activity. Bhattacharyya *et al.* showed the presence of PCs and lysoPCs with very long fatty acids significantly decreased in overdose conditions indicative of a structure-activity relation with enzymes responsible for phospholipid metabolism.[30] Recently, Yamada *et al.* showed mechanisms where acute liver failure induced by APAP toxicity is ferroptosis-mediated which is driven by polyunsaturated fatty acids.[31] Ming *et al.* have also reported that the APAP overdose conditions induce dramatic changes to PC and PE profiles of plasma and liver through possible hepatocyte damage and interferences to phospholipid metabolism.[29] Therefore, the importance of investigating APAP effects on pure phospholipid bilayer structures becomes an important research question from a fundamental, material, and chemical standpoint.

Here we have investigated the structural and dynamic details of drug-lipid vesicle interactions in hope of broadening the knowledge of biophysical processes that may carry links to many long-standing questions in the realm of the drug's side-effects and toxicity effects at the molecular basis. We hypothesized the contribution from incorporated Acetaminophen to the phospholipid bilayer is enough to contribute to changes in physicochemical properties of the bilayer, the morphology of the vesicle self-assembly as well as lipid dynamics at nanometer length-scale and nanosecond timescale which will result in functional changes to the membrane-bound enzymes, etc. regulating Acetaminophen mechanisms of action and its side/overdose effects. To investigate these structural, morphological, and dynamical changes we used DOPC large unilamellar vesicles with APAP (Figure 1) and conducted dynamic light scattering (DLS), small-angle X-ray and neutron scattering (SAXS/SANS), cryo-transmission electron microscopy (cryo-TEM), and neutron spin-echo spectroscopy studies (NSE).

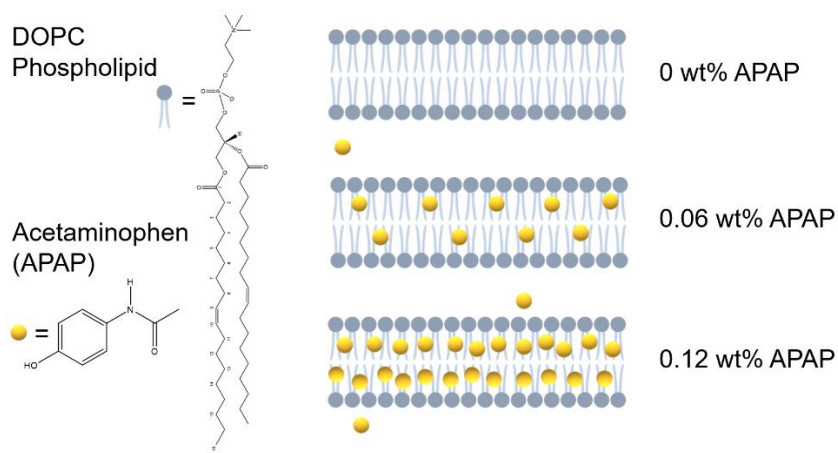

Figure 1. Chemical structures of DOPC phospholipid and Acetaminophen (APAP) on the left and the schematic representation of the 3 scenarios of gradual incorporation of APAP at 0 wt%, 0.06 wt%, and 0.12 wt% explored on the right.

## Theoretical Background

**Vesicle structure:** The vesicle form factor is modeled using an extension of the core-shell model used in our previous studies.[32-36] For unilamellar vesicles the core-shell model consists of a water core of radius $R_c$, encapsulated by three shells with (i) lipid inner-head, (ii) tail region (hydrocarbon core), and (iii) outer-head. The 1D scattering pattern is given by:

$$P(Q, R, t, \Delta\rho) = \frac{\phi A^2(Q)}{V(r_3) - V(R_c)} \quad (1)$$

with $\phi$ is the lipid volume fraction. The scattering contribution from three different shells are given by

$$A^2(Q) = A_1^2 + A_2^2 + A_3^2 + A_{12} + A_{23} + A_{13} \quad (2)$$

with $A_{12} = A_1 A_2$, $A_{12} = A_2 A_3$, and $A_{13} = A_1 A_3$ are the cross-terms for inner-head-tail, tail-outer-head and the two outer lipid head layers, respectively.

For a spherical Bessel's function $j_1(x) = \frac{\sin(x) - x\cos(x)}{x^2}$, and volume $V(r) = \frac{4}{3}\pi r^3$ we have individual scattering amplitudes as:

$$
\begin{aligned}
A_1^2 &= (\rho_{head} - \rho_{solv})^2 \left[ 3V(r_1)\frac{j_1(Qr_1)}{Qr_1} - 3V(R_c)\frac{j_1(QR_c)}{QR_c} \right]^2 \\
A_2^2 &= (\rho_{tail} - \rho_{solv})^2 \left[ 3V(r_2)\frac{j_1(Qr_2)}{Qr_2} - 3V(r_1)\frac{j_1(Qr_1)}{Qr_1} \right]^2 \\
A_3^2 &= (\rho_{head} - \rho_{solv})^2 \left[ 3V(r_3)\frac{j_1(Qr_3)}{Qr_3} - 3V(r_2)\frac{j_1(Qr_2)}{Qr_2} \right]^2 \\
A_{12} &= 2(\rho_{tail} - \rho_{solv})(\rho_{head} - \rho_{solv}) \left[ 3V(r_1)\frac{j_1(Qr_1)}{Qr_1} - 3V(R_c)\frac{j_1(QR_c)}{QR_c} \right] \left[ 3V(r_2)\frac{j_1(Qr_2)}{Qr_2} \right. \\
&\quad \left. - 3V(r_1)\frac{j_1(Qr_1)}{Qr_1} \right] \\
A_{23} &= 2(\rho_{tail} - \rho_{solv})(\rho_{head} - \rho_{solv}) \left[ 3V(r_2)\frac{j_1(Qr_2)}{Qr_2} - 3V(r_1)\frac{j_1(Qr_1)}{Qr_1} \right] \left[ 3V(r_3)\frac{j_1(Qr_3)}{Qr_3} \right. \\
&\quad \left. - 3V(r_2)\frac{j_1(Qr_2)}{Qr_2} \right] \\
A_{13} &= 2(\rho_{head} - \rho_{solv})(\rho_{head} - \rho_{solv}) \left[ 3V(r_1)\frac{j_1(Qr_1)}{Qr_1} - 3V(R_c)\frac{j_1(QR_c)}{QR_c} \right] \left[ 3V(r_3)\frac{j_1(Qr_3)}{Qr_3} \right. \\
&\quad \left. - 3V(r_2)\frac{j_1(Qr_2)}{Qr_2} \right]
\end{aligned}
\quad (3)
$$

Here, $r_1 = R_c + t_{head}$, $r_2 = R_c + t_{head} + t_{tail}$, $r_3 = R_c + 2t_{head} + t_{tail}$. The membrane thickness of the bilayer from SANS is given by, $\delta_{HH}(SANS) = 2t_{head} + t_{tail}$. For DOPC we used the neutron scattering length density (NSLD) of the hydrocarbon tail, $\rho_{tail} = -2.08 \times 10^9$ cm$^{-2}$, and for phosphatidylcholine (PC) head group, $\rho_{head} = 1.73 \times 10^{10}$ cm$^{-2}$ has been used.[37] For D$_2$O the NSLD of the solvent of, $\rho_{solv} = 6.36 \times 10^{10}$ cm$^{-2}$ has been used. Each shell thickness and scattering length density is assumed to be constant for the respective shell. The macroscopic scattering cross-section is obtained by

$$\frac{d\Sigma}{d\Omega}(Q)_{SANS} = \int dr P(Q, R, t, \Delta\rho) s(r) \tag{4}$$

For the size polydispersity, $s(r)$, we used a log-normal distribution.[38]

A Gaussian distribution was used to include polydispersity of the bilayer and the water layer in the case of multilamellar vesicles.

**Bilayer structure**: To get direct access to the macroscopic scattering cross-section given by the SAXS scattering intensity from random lamellar sheet consisting of lipid heads and tails of thicknesses are $\delta_H$ and $\delta_T$, respectively is given by

$$\frac{d\Sigma}{d\Omega}(Q)_{SAXS} = 2\pi \frac{\phi P(Q) S(Q)}{Q^2 d} \tag{5}$$

with particle volume fraction, $\phi$, and the lamellar repeat distance, $d$. The form factor is given by:

$$P(Q) = \frac{4}{Q^2} \left[ \Delta\rho_H \{\sin(Q(\delta_H + \delta_T)) - \sin(Q\delta_T)\} + \Delta\rho_T \sin(Q\delta_T) \right]^2 \tag{6}$$

The scattering contrasts for the head and tail are $\Delta\rho_H$ and $\Delta\rho_T$, respectively. The corresponding thicknesses are $\delta_H$ and $\delta_T$, respectively. The head to head bilayer thickness is given by, $\delta_{HH}(SAXS) = 2(\delta_H + \delta_T)$. For unilamellar structure, $N = 1$, the repeat distance in equation 6 is given by the membrane thickness, $\delta_{HH}$. The Caille structure factor is given by

$$S(Q) = 1 + 2 \sum_{n=1}^{N-1} \left(1 - \frac{n}{N}\right) \cos(Qdn) \exp\left(-\frac{2Q^2 d^2 \alpha(n)}{2}\right) \tag{7}$$

with the number of lamellar plates, $N$ ($N > 1$), and the correlation function for the lamellae, $\alpha(n)$, defined by

$$\alpha(n) = \frac{\eta_{cp}}{4\pi^2} (\ln(\pi n) + \gamma_E) \tag{8}$$

with $\gamma_E = 0.57721$ the Euler's constant. The elastic constant for the membranes is expressed in terms of the Caille parameter, $\eta_{cp} = \frac{Q_1^2 k_B T}{8\pi \sqrt{(\kappa_c \bar{B})/\delta_{HH}}}$, where $\kappa_c$ and $\bar{B}$ are the bending elasticity and the compression modulus of the membranes. Here $\bar{B}$ is associated with the interactions between the membranes. The position of the first-order Bragg peak is given by $Q_1$, whereas $k_B$ is the Boltzmann's constant and $T$ the absolute temperature. A Gaussian distribution function includes thickness polydispersity for $d$, $\delta_H$ and $\delta_T$.

**Dynamics of a lipid bilayer**: The overall membrane dynamics can be a result of the superposition of different motions over different length and time scales. Neutron Spin Echo (NSE) spectroscopy has been vital to understanding the membrane fluctuations at the length scale of the lipid-bilayer thickness.[33, 35, 39-42] The dynamic structure factor or the intermediate scattering function, $S(Q,t)$, from NSE can be modeled assuming statistically independent tail-motion, height-height fluctuations (membrane undulations) and translational diffusion (of the entire vesicle) [39]

$$S_{liposome}(Q,t) = \left( n_{H,head} + n_{H,tail} \left( \mathcal{A}(Q) + (1 - \mathcal{A}(Q)) \exp\left(-\left(\frac{t}{\tau}\right)^{\beta}\right) \right) \right) S_{ZG}(Q,t) \exp(-D_t Q^2 t) \quad (9)$$

Here the relative fractions of protons in the head are kept fixed to, $n_{H,head} = 0.21$, for h-DOPC, and, $n_{H,tail} = 1 - n_{H,head} = 0.79$. The variable $\mathcal{A}(Q)$ represents the elastic fraction of the lipid tail motion. The exponential term represents the membrane undulation within the Zilman-Granek (ZG) model [43]

$$S_{ZG}(Q,t) \propto \exp\left[-(\Gamma_Q t)^{2/3}\right]. \quad (10)$$

The Q-dependent decay rate, $\Gamma_Q$, can be used to determine the intrinsic bending modulus, $\kappa_\eta$, by[44-46]

$$\frac{\Gamma_q}{Q^3} = 0.0069 \gamma \frac{k_B T}{\eta} \sqrt{\frac{k_B T}{\kappa_\eta}} \quad (11)$$

Here $\eta$ is the viscosity, $k_B$ the Boltzmann constant, $T$ the temperature. For lipid bilayers, $\gamma = 1$.[41-43, 45-47]

To obtain model-independent data we analyze the mean squared displacement ($\langle \Delta r(t)^2 \rangle$, MSD), and the non-Gaussianity parameter, $\alpha_2(t) = \frac{d}{d+2} \frac{\langle \Delta r(t)^4 \rangle}{\langle \Delta r(t)^2 \rangle^2} - 1$, from the measured dynamic structure factor, $S(Q,t)$, using a cumulant expansion given by, [33, 39, 48, 49]

$$\frac{S(Q,t)}{S(Q)} = A \exp\left[-\frac{Q^2 \langle \Delta r(t)^2 \rangle}{6} + \frac{Q^4 \alpha_2(t)}{72} \langle \Delta r(t)^2 \rangle^2\right] \quad (12)$$

The non-Gaussianity parameter $\alpha_2$ is essentially defined as a function of the fourth $\langle \Delta r(t)^4 \rangle$ and the second moment squared $\langle \Delta r(t)^2 \rangle^2$, Here we use the space dimension, $d = 3$.[33, 49, 50]

An alternative representation of equations 10 and 11 is[39]

$$\frac{\kappa_\eta}{k_B T} = \frac{t^2}{c(\eta,T)^3 \langle \Delta r(t)^2 \rangle^3} \tag{13}$$

with $c(\eta,T) = \frac{1}{6}\left(\frac{\eta}{0.0069 k_B T}\right)^{2/3}$. Equation 13 is strictly valid within the ZG approximation $\langle \Delta r(t)^2 \rangle \propto t^{2/3}$. Any deviation indicates dynamics beyond ZG undulations

## Experimental

**Materials** All chemicals and reagents were used as received. 1,2-di-(octadecenoyl)-*sn*-glycero-3-phosphocholine (DOPC) (MC 8181, Lot No. 1905811L) was purchased from NOF America Corporation (White Plains, NY, USA). Acetaminophen (APAP) powder, USP (Lot No. 2HE0496) was purchased from Spectrum Chemical MFG CORP (New Brunswick, NY, USA). Acetaminophen powder was kept in the freezer (-20 °C) in dry-dark conditions until use due to its light sensitivity. All other organic reagents (HPLC grade) and $D_2O$ (99.8% deuterated) were received from Sigma Aldrich (St. Louis, MO, USA).

**Sample preparation** First, DOPC lyophilized powder was dissolved in chloroform (40 mg/mL) and Acetaminophen in isopropanol (4.8 mg/mL). Then, DOPC and Acetaminophen dissolved in respective organic solvents were mixed to obtain specific molar/weight ratios and were dried using a rotary evaporator and vacuum overnight to remove the solvent traces. The obtained dry lipid cakes were hydrated with $D_2O$. The vesicle suspensions were then subjected to freeze-thaw cycling (-20 °C/50°C) in ten-minute intervals. Finally, the vesicles were extruded using an Avanti mini-extruder with 100 nm polycarbonate membranes passing the vesicle suspension 33 times through the membrane to obtain large unilamellar vesicles ~ 100 nm. The final DOPC concentrations of the samples were 1 wt% with the APAP concentrations being 0 wt% (pure DOPC), 0.06 wt%, and 0.12 wt% respectively. Molar ratios of DOPC: APAP for the samples are (1:0, 3:1, and 2:3).

All measurements were conducted at 1 wt% DOPC concentration at ambient temperature 20 - 25 °C where DOPC lipid is in the fluid phase. Results shown in this MS are for two different sample preparations due to the different dates of accessibility to respective instruments. (Batch 1: DLS, SANS, SAXS (at Stanford Synchrotron) and NSE) (Batch 2: cryo-TEM and SAXS (at LSU-CAMD) Samples and data were reproducible within experimental accuracy (See SI).

**Dynamic Light Scattering (DLS)** DLS experiments were conducted at LSU Polymer Analysis Laboratory (LSU PAL) using a Malvern Zetasizer Nano ZS instrument. (Specifications: He-Ne laser wavelength, $\lambda = 633$ nm at 30 mW laser power, backscattering set up angle $\theta = 173°$). A 0.5 mL aliquot of each vesicle suspension was added to a disposable micro-UV cuvette and the respective DLS data were obtained using $D_2O$ as the solvent at 25 °C. The hydrodynamic radius, $R_h$, of the liposomes in pure DOPC vesicles and each of the DOPC/Acetaminophen vesicles was calculated from the translational diffusion coefficient, $D_t$, using the Stokes-Einstein equation, $R_h = k_B T/(6\pi\eta_0 D_t)$, with the Boltzmann constant, $k_B$, the temperature, $T$, the viscosity of the solvent, $\eta_0$. DLS measurements were triplicated for each sample and averaged. The results are listed in Table 1.

**Cryo Transmission Electron Microscopy (cryo-TEM)** The cryo-TEM data were obtained at the BioCryo facility of Northwestern University Nuance Center (remote access due to COVID19 restrictions). The samples were applied to 200 mesh Cu grids with a lacey carbon membrane. Before plunge-freezing, grids (EMS Cat# LC200-CU-100) were glow discharged in a Pelco easiGlow glow discharger (Ted Pella Inc., Redding, CA, USA) using an atmosphere plasma generated at 15 mA for 15 seconds with a pressure of 0.24 mbar. This treatment created a negative charge on the carbon membrane, allowing liquid samples to spread evenly over the grid. A 4 µL volume of each sample was pipetted onto the grid and blotted for 5 seconds with a blot offset of +0.5 mm, followed by immediate plunging into liquid ethane within an FEI Vitrobot Mark III plunge freezing instrument (Thermo Fisher Scientific, Waltham, MA, USA). Grids were then transferred to liquid nitrogen for storage. The plunge-frozen grids were kept vitreous at –180 °C in a Gatan Cryo Transfer Holder model 626.6 (Gatan Inc., Pleasanton, CA, USA) while viewing in a Hitachi HT7700 W-emission TEM at 100 kV. Image data were collected by a Gatan Orius $4 \text{ k} \times 2.67 \text{ k}$ digital camera (Gatan Inc., Pleasanton, CA, USA). Samples required no dilution at 1 wt% concentration for successful visualization. Cryo-TEM images were analyzed using ImageJ software.

**Small-Angle X-ray Scattering (SAXS)**

SAXS data were obtained on two occasions. SAXS results displayed in Figure 3 were obtained at the Stanford Synchrotron Radiation Light source (SSRL), beamline 4-2.[51] An automated sampler system was used to load 40 µL aliquots of samples to the capillary cell where the sample was exposed to the X-ray radiation 24 times in 1 s exposures as the flow cell gently moves back and forth to minimize any potential radiation damage by continuous spot exposure. The scans were then statistically averaged, and radial averaging over the same data yield the intensity as a function of the momentum transfer. The data were obtained at a beam energy of 11 keV, at the detector distance of 1 m to explore a $Q$ range of 0.01 – 1 Å$^{-1}$ using a Pilatus $3 \times 1$ M detector. Data reduction was done using standard SSRL protocols implemented in the software Blu-Ice.[52] SAXS data obtained at Louisiana State University Center for Advanced Microstructures and Devices (CAMD) are listed in the SI. The CAMD SAXS/WAXS/GISAXS beamline at LSU, Baton Rouge, Louisiana was manufactured by Saxslab, now Xenocs (France, USA) by including the preexisting sample chamber of the former SAXS beamline. The experiments were performed in the lab mode with a Genix Cu-K$_\alpha$ lab source (Xenocs). A Pilatus 3R 300K detector was placed outside the evacuated flight tube at a sample to detector distance of 263.40 mm. In order to avoid additional windows in the flight path, the sample chamber was set under mild vacuum. The sample was placed inside a borosilicate glass capillary (Hilgenberg), with a diameter of 1 mm Data reduction was performed with the SAXSGUI program.

**Small-Angle Neutron Scattering (SANS)** SANS data were obtained at NG-7 SANS instrument at National Institute of Standards and Technology (NIST), NIST Center for Neutron Research (NCNR).[53] The sample-to-detector distances, $L_{sd}$, were set to 1, 4, and 13 m, at neutron wavelength, $\lambda = 6$ Å. Another configuration with lenses at $L_{sd} = 15.3$ m, and $\lambda = 8$ Å was used to access low $Q$'s[54] covering a total $Q$-range from 0.001 to 0.6 Å$^{-1}$, where $Q = 4\pi \sin(\theta/2)/\lambda$, with the scattering angle, $\theta$. A wavelength resolution of, $\Delta\lambda/\lambda = 14\%$, was used. All data reduction into intensity, $I(Q)$, vs. momentum transfer, $Q = |\vec{Q}|$, was carried out following the standard procedures that are implemented in the NCNR macros to the Igor software package.[55] The intensity values were scaled into absolute units (cm$^{-1}$) using direct beam. The D$_2$O as the solvent and empty cell were measured separately. D$_2$O was subtracted as buffer background as well as empty cell scattering before analysis. Error bars are expressed as one standard deviation.

**Neutron Spin-Echo Spectroscopy (NSE)** NSE measurements were obtained at BL15 instrument at the Spallation Neutron Source of the Oak Ridge National Laboratory, Oak Ridge, TN.[56] Hellma quartz cells with 2 mm sample thickness were used to mount the samples. Measurements were conducted using a neutron wavelength of 8 Å. The BL15 ECHODET software package was used for data reduction. Two background samples (D$_2$O and Acetaminophen saturated D$_2$O) were measured separately and used for background subtraction accordingly.

## Results and Discussion

First, the structure was investigated using a combination of DLS, cryo-TEM, SANS, and SAXS. DLS results showed a slight decrease in the hydrodynamic radius of the vesicles with increasing Acetaminophen (APAP) concentration in the lipid bilayer, Table 1. Size distribution increased with the addition of APAP, (Figure 2.a), as reflected by the polydispersity of the vesicles calculated by log-normal distribution fits, Table 1. SANS data were modeled (Figure 2b) using a spherical core-shell model as described in the Theoretical Background section. SANS data showed no significant change of lipid head-to-head distance of the bilayer ($\delta_{HH}$) which remained at approximately 4 nm. However, as obtained from the aggregation number, $N_{agg} = V_s/V_l$, illustrates the decrease in the number of lipids per liposome with increasing APAP concentration. Here, $V_s$ is the shell volume of the liposome and $V_l$ the molar volume of the phospholipid. SANS appears to be more sensitive in obtaining the reduction in the size of the liposomes than DLS. This might be due to the increase in polydispersity from DLS. The observed polydispersity can be due to the size or shape polydispersity. In this case it accounts for the shape polydispersity as seen from cryo-TEM images (Figure 4). Cryo-TEM data demonstrate heterogeneous morphology of liposomes appears to increase with higher concentrations of APAP. The effects of APAP on lipid vesicle deformation are especially apparent in the saturated APAP solution (Figure S4). In SANS, the data is modeled using the statically dominant spherical core-shell particle with uniform polydispersity for different APAP concentrations. Comparing results from DLS and SANS suggests that although the spherical nature of the particles is statistically dominant, we might have additional particles of different shapes causing an increase in DLS polydispersity.

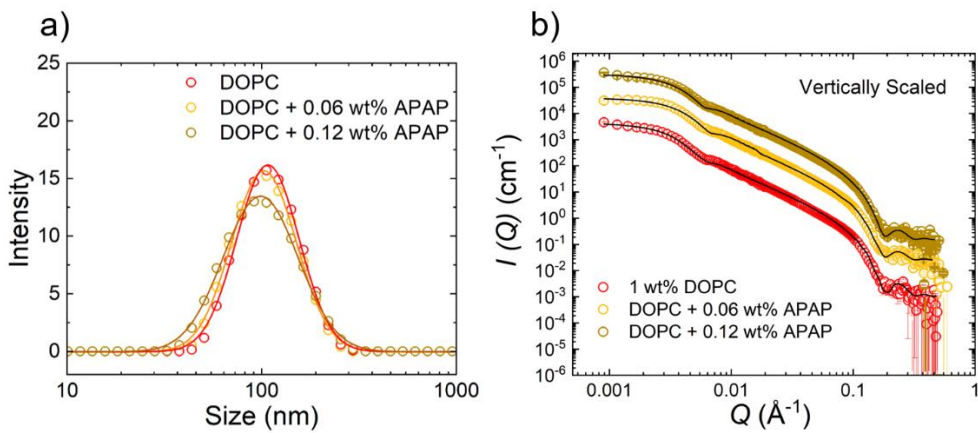

Figure 2. a) DLS intensity size distribution of 1wt% DOPC vesicles (red-empty) and DOPC with APAP (Acetaminophen) with 0.06 wt% (yellow-empty) and 0.12 wt% (yellow green-empty). The solid lines show log-normal fits. b) SANS curves for the same DOPC vesicles and DOPC/APAP vesicles. Black lines show vesicle model fits. Data are vertically scaled for better visualization. The size and other parameters obtained are listed in Table 1.

Table 1. Parameters found from SANS and DLS in Figure 2. $R_{SANS}$ vesicle radius from SANS, $\delta_{HH}$ is lipid head-to-head distance, $N_{agg}$ is aggregation number depicting the number of lipids per liposome, PD is polydispersity and Size DLS, $R_h$ is the hydrodynamic radius of the vesicles found from the log-normal fitting of the size distribution.

| Acetaminophen Concentration (wt%) | $R_{SANS}$ (nm) | $\delta_{HH}$ (nm) | $N_{agg} \times 10^4$ | $PD_{SANS}$% | $R_h$ (nm) DLS | $PD_{DLS}$ % |
|---|---|---|---|---|---|---|
| 0.00 | 51.2 ± 0.4 | 3.9 ± 0.1 | 9.3 ± 0.2 | 27 | 61.6 ± 0.2 | 37 |
| 0.06 | 43.5 ± 0.6 | 4.0 ± 0.3 | 6.7 ± 0.7 | 27 | 59.5 ± 0.2 | 38 |
| 0.12 | 46.4 ± 0.5 | 3.9 ± 0.2 | 7.5 ± 0.5 | 27 | 60.1 ± 0.3 | 44 |

While SANS provides information on the liposome diameter, due to its resolution small-angle X-ray scattering (SAXS) can be used to obtain an accurate estimate of the lipid head and tail thickness as well as the number of bilayers. This sensitivity is achieved due to the reduced influence from vesicle form-factor over the relevant $Q$-range, higher X-ray contrast for phosphorous atoms in the lipid heads, and excellent instrumental resolution. Figure 3 illustrates SAXS diffraction patterns for 1 wt% DOPC with 0 to 0.12 wt% Acetaminophen concentrations.

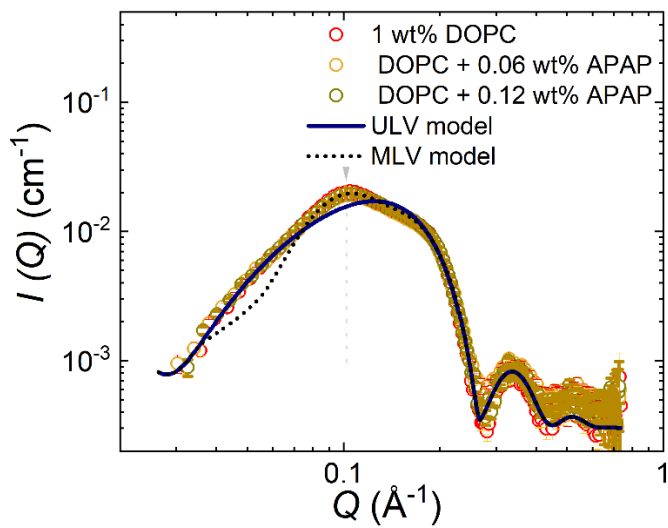

Figure 3. SAXS scattering data DOPC vesicles and DOPC/APAP vesicles in logarithmic scales. The data is modeled for ULV ($N = 1$) (blue line) and a MLV model ($N = 2$) (black dashed line) respectively.

We observed only a negligible difference in the scattering pattern for different Acetaminophen concentrations. The peak at $Q = 0.1\,\text{Å}^{-1}$ (vertical arrow) indicates a small deviation from the simple unilamellar vesicle model (ULV model) (blue solid line), with $N = 1$ in equation 5. This peak corresponds to the evolution of the inter-bilayer structure factor peak and is modeled using a multilamellar vesicle model (MLV model) (black dashed line) described by equations 5 and 6 for $N = 2$ layers, where $\eta_{\text{cp}}$ is kept constant at 0.01. The first-order Bragg peak is given by $Q_1 = 0.1\,\text{Å}^{-1}$ corresponds to a lamellar repeat distance $d = 6.3\,\text{nm}$ and the average value of the bilayer thickness $\delta_{HH} = 4.14 \pm 0.5\,\text{nm}$, within experimental accuracy it is the same for different Acetaminophen concentrations. A Gaussian polydispersity of $22\,\%$ was introduced to model $d$. The small discrepancy for $Q < 0.07\,\text{Å}^{-1}$ is attributed to the presence of a mixture of ULV, MLV, and overlapping vesicle structures, resulting in an apparent higher polydispersity than expected for a simple MLV model.[57] Such a scenario can occur due to a small percentage of MLVs present in the vesicle suspension and which is evident by some of the cryo-TEM images, Figure 4 and Supporting Information (SI).

In order to visualize the morphology of the vesicles at different Acetaminophen concentrations, cryo-TEM was used. Cryo-TEM enables the visualization of phospholipid vesicles in their native state.[58] As shown below in Figure 4a), pure DOPC vesicles have a spherical morphology. The presence of occasional bilamellar vesicles confirms the slight deviation of bilayer structure from the ULV model in SAXS analysis which was previously discussed. From the particles counted in cryo-TEM images, over 76% were unilamellar vesicles. When it comes to the presence of Acetaminophen, deviations from the original spherical shape of the vesicles can be observed in Figures 4b) and 4c).

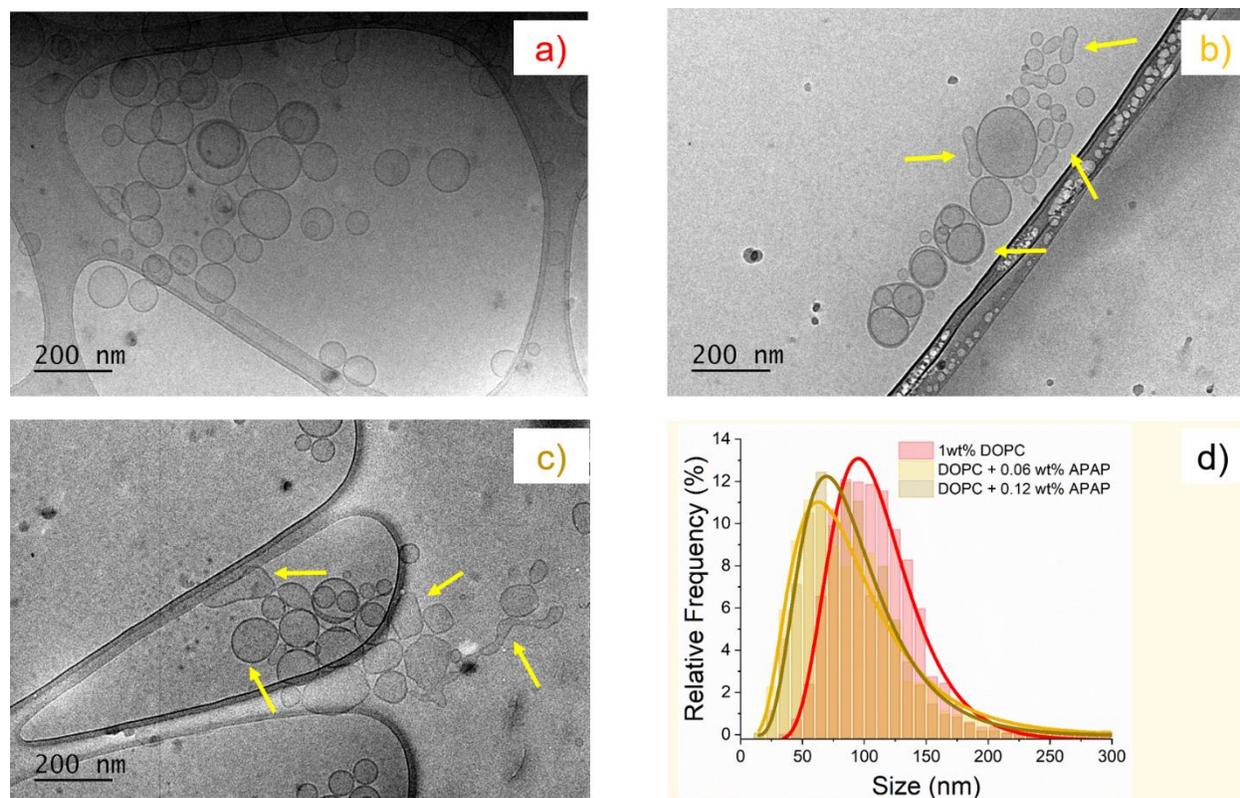

Figure 4. cryo-TEM images and analysis of DOPC vesicles with Acetaminophen. a) 1wt% Pure DOPC in $D_2O$, b) 1wt% DOPC with 0.06 wt% APAP, c) 1wt% DOPC with 0.12 wt% APAP and d) Size distributions (particle analysis) for a, b, and c samples (red, yellow, and yellow-green respectively). The yellow arrows indicate vesicles departing from the original spherical state. More Images and analysis details are included in the SI.

Vesicles appeared more fluid-like and had irregular sphere shapes as well as occasional tubular shapes indicated by the yellow arrows. Such deviations appear to be more pronounced when the Acetaminophen concentration was increased from 0.06 wt% to 0.12 wt%. Furthermore, overall vesicle sizes decreased in the presence of Acetaminophen. Figure 4d) shows the size distribution of vesicles in each sample measured using multiple cryo-TEM images. More than 1500 size measurements were used for the size analysis using the log-normal distribution model to describe data. The trend of size variation was similar to DLS and SANS results previously presented in Table 1. The polydispersity of the particles increased agreeing with results obtained by DLS and SANS. More images and details of the analysis are provided in the SI.

Membrane dynamics were measured by neutron spin-echo (NSE) spectroscopy. Data were modeled by the multiplicative model (equation 9) which includes vesicle translational diffusion, membrane fluctuations, and confined motion of lipid tails for dynamic structure factor $S(Q,t)$ analysis.[39] All experiments were conducted at room temperature. Therefore, DOPC lipids are in the fluid phase. The model agreed well with the data in the $Q$ and Fourier time range used for the experiment (Figure 5). It should be noted that within the NSE time window the translational diffusion, $D_t$, is almost by a factor 5 slower than the ZG decay as observed from the dynamic structure factor illustrated in Figure S7 in SI. The details of the NSE data analysis are presented in the SI.

Bending rigidity values, $\kappa_\eta$, calculated from the analysis are displayed in Table 2. Briefly, for 1wt% DOPC, and 1 wt% DOPC with 0.06 wt% and 0.12 wt% Acetaminophen, we obtained membrane rigidity values: $17.0 \pm 2.0$, $10.3 \pm 1.0$ and $9.0 \pm 1.0$ $k_B T$ respectively. It should be noted that the small effects of MLV and fused vesicle structures as observed from SAXS and cryo-TEM images have been neglected in the calculation of the membrane rigidity from equation 11.

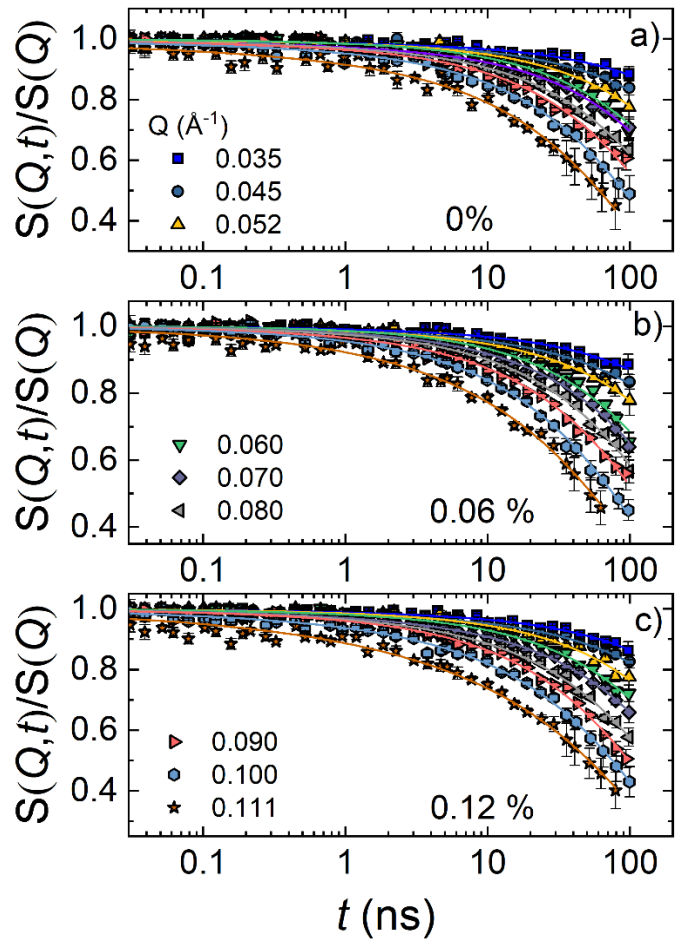

Figure 5. Dynamic structure factor, $S(Q,t)/S(Q)$, as a function of Fourier time, $t$, for different $Q$'s, for a) pure 1wt% DOPC (0 wt% Acetaminophen), b) with 0.06 wt% Acetaminophen (APAP) and c) DOPC with 0.12 wt% APAP at Room Temperature. The data are analyzed using multiplicative model[39]

Figure 6 expressing the dynamics using the MSD, which provides information without relying on a specific model. The corresponding changes in the slope in the double-logarithmic plot or the respective power laws indicate different dynamical processes.

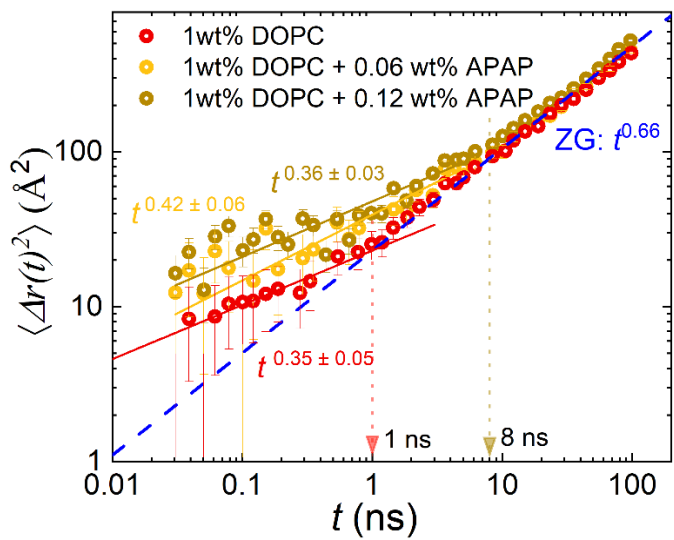

Figure 6. Mean square displacement, $\langle \Delta r(t)^2 \rangle$, vs. Fourier time, t, for 1 wt% DOPC (red) and DOPC with Acetaminophen (APAP): APAP 0.06 wt% (yellow), DOPC with APAP 0.12 wt% (yellow-green). The solid lines represent the experimental power-law dependence for each sample. The blue dashed line shows the ZG $t^{0.66}$ for reference. Vertical dashed lines at 1 ns and 8 ns show deviation from ZG behavior below ~1 ns for pure DOPC vesicles and ~8 ns for vesicles with APAP respectively.

Within the time window of our NSE experiment, the MSD highlights two different time domains as indicated by different power laws. (1) Above 8 ns in Fourier time, results point to $t^{0.66}$ dependence for all samples. Such an exponent point to a ZG region corresponding to membrane fluctuations. With the addition of APAP, the value of $\langle \Delta r(t)^2 \rangle$ increases with the concentration, which would be consistent with a stronger fluctuation that leads to a lower bending modulus, compatible with the parameters obtained by the fit of the multiplicative model. Data representation in the double logarithmic plot hides that the MSD in this region changes by a factor of 1.03 (0.06 wt% APAP) or 1.13 (0.12 wt% APAP) compared to the pure DOPC (SI Figure S9a). (2) For low Fourier times, DOPC vesicles show a different power law, with a slope and onset that seem to change with the APAP concentration. However, within the statistical accuracy of the experimental NSE data, the results would also be compatible with an unchanged slope. This explanation is supported by our previous results on DOPC, which showed a power-law exponent $0.26 \pm 0.03$.[33,40] Considering the standard deviation the results agree but point to caution when it comes to the deduction of a systematic change with the APAP concentration. Instead, the systematic change of the MSD with increasing APAP concentration in the short time regions seems to be statistically more significant than the change in the ZG range. In this region, MSD changes by a factor of 1.45 (0.06 wt% APAP) or 1.75 (0.12 wt% APAP) compared to the pure DOPC (SI Figure S9b). This change points to more spatial freedom for the tail motion once APAP has been added to the system. The transition to the ZG region seems to depend on the concentration of the APAP as well.

Considering the fast dynamics at $t < 8$ ns, we have calculated $\kappa_\eta/k_BT$ using equation 11 as a function of the Fourier time over the entire NSE time window. The results for 0, 0.06% and 0.12% acetaminophen concentrations are presented in Figure 7. The results are calculated using $\eta = \eta_{solvent}$ and for $D_t = 0$.[33, 39, 40] The average bending rigidity values found from the analysis within the ZG region are displayed in Figure 7 and are similar to that observed from the multiplicative model (equation 9) as displayed in Table 2.

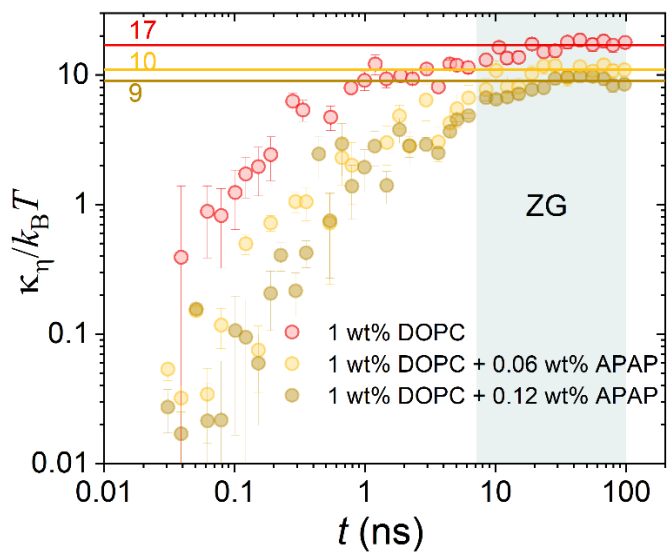

Figure 7. The membrane rigidity, $\kappa_\eta$, divided by thermal energy, $k_BT$, with the Boltzmann constant, $k_B$, and the temperature, $T$, as a function of Fourier time $t$. The data is calculated over the NSE time window from the MSD, for 1wt% DOPC and DOPC with Acetaminophen (0.06 wt% and 0.12 wt%) at Room Temperature. The calculated average values from the flat ZG region are illustrated by the horizontal lines. These lines represent the bending modulus, $\kappa/k_BT$. Values are listed for visualization.

Table 2. Membrane bending rigidity of 1wt% DOPC vesicles and DOPC/APAP vesicles from NSE obtained from the analysis presented in Figure 7.

| 1wt% DOPC vesicles with Acetaminophen concentration wt% | Bending modulus $\kappa_\eta$ ($k_BT$) | Normalized bending rigidity $\dfrac{\kappa_\eta(DOPC + APAP)}{\kappa_\eta(DOPC)}$ |
|---|---|---|
| 0 | 17.0 ± 2.0 | 1.0 |
| 0.06 | 10.3 ± 1.0 | 0.6 |
| 0.12 | 9.0 ± 1.0 | 0.5 |

The NSE experiments show that the incorporation of Acetaminophen in lipid bilayer has decreased the membrane rigidity of DOPC vesicles significantly. Comparing our results to existing experiments that have investigated membrane rigidity changes with other drugs such as Ibuprofen, Aspirin, Indomethacin, etc. previously, the decrease in membrane rigidity is comparable. To obtain model-independent insights, we have compared the normalized membrane rigidities. In order to see a trend with molar fraction of drugs in the lipid membrane, and decrease in membrane rigidity, the $\kappa_\eta$ the ratio is obtained by normalizing compared to the pure phospholipid vesicle systems used in the respective studies (Table 3).

Table 3. Comparison of normalized bending rigidity values of phospholipid vesicles in the fluid phase with various small drug molecules as a function of the molar percentage of the drugs.

| Drug | Lipid and Conditions | Molar % Drug | Method | Normalized Bending Rigidity $\frac{\kappa_\eta(Drug:Lipid)}{\kappa_\eta(Pure\ Lipid)}$ | Reference |
|---|---|---|---|---|---|
| No Drugs | DOPC, LUV | 0 | NSE-multiplicative model | 1 | *This work* |
| Acetaminophen | DOPC, LUV | 25 | NSE-multiplicative model | 0.6 | *This work* |
| Acetaminophen | DOPC, LUV | 60 | NSE-multiplicative model | 0.53 | *This work* |
| Ibuprofen | DMPC | 25 | NSE- ZG model | 0.67 | [59] |
| Indomethacin | DMPC | 25 | NSE- ZG model | 0.61 | [59] |
| Aspirin | DMPC | 25 | NSE- ZG model | 0.81 | [59] |
| Aspirin | DMPC | 36 | NSE- ZG model | 0.67 | [60] |
| Salicylate | SOPC, GUVs | N/A (1 mM) | Micropipette aspiration and dynamic tension spectroscopy | 0.67 | [19] |
| Ibuprofen | DMPC (pH < 2), SUV | 30 | NSE- ZG model | 0.67 | [20] |
| Ibuprofen | DMPC (pH > 2), SUV | 30 |  | 0.47 | [20] |

Although the results represent different phospholipids, experimental conditions, and techniques, all phospholipids are at conditions where they are in the fluid phase, making the comparison valid. A general trend of decreasing membrane rigidity with the incorporated molar percentage of the drugs can be observed. However, the severity of impact is different for drugs explored showing that each drug can have unique impacts on the lipid membrane dynamics.

## Conclusions

In summary, we have investigated the concentration-dependent impact of Acetaminophen (APAP) or commonly known as paracetamol on DOPC LUVs in the fluid phase. We observed a slight decrease in vesicle size by DLS and SANS. Our cryo-TEM experiments illustrated further morphological changes. Vesicle shapes became more and more irregular with the increasing concentration of Acetaminophen which might be an explanation of the increasing polydispersity of the vesicle populations observed by DLS. In SAXS, surprisingly, we did not observe much change in the bilayer structure. Previous studies have shown substantial differences in SAXS profiles when small drug molecules alter the melting temperature $T_m$ of the lipid bilayers.[61] We assume that the relatively small changes in the SAXS profile may relate to the fluid phase (experiments are conducted well above the $T_m$ of DOPC).

We observe a change in aggregation number and find a decrease by almost 28% (0.06 wt% APAP) and 19% (0.12 wt% APAP) compared to the pure DOPC (0 wt% APAP). At the same time, we find a decrease in the diameter, which leads to the surprising effect that the radius ascribed to a lipid stays almost constant within the experimental accuracy. Eventually, a slight change of 1.06 can be calculated. Surprisingly, simulations on the system APAP and DPPC, and DMPC revealed a virtually constant value.[27] Though, simulations and our experiments are different both results point to a little to no effect on the area per lipid.

Surprisingly, we observe a stronger change of the mean square displacement of the lipid tail motion. It seems that the space explored by the lipid tails increases by a factor of 1.45 (0.06%), and 1.75 (0.12 %) compared to DOPC without the drug. At the first glance, one would think that the samples have been different. However, the simulations by Nademi et al. also observed a virtually constant area per lipid but a change of the lateral diffusion. In the case of Nademi, the authors calculated the mean square displacement of the lipid while our experiments see mainly the lipid tail motion. (Neutrons measure the protons, and most of the protons are in the tail.) The simulations find a substantial decrease, which the authors ascribe to drug molecules entering the space between the lipids. Our results point also to drug molecules increasing the space between lipids, which would explain the larger MSD of the tails. As the distance between the molecules determines the molecular interactions [Nademi] the reduced bending modulus seems to be a direct consequence of the increased spatial freedom. Again, we point to the surprising area per lipid from the SANS experiments which is virtually unchanged (at most changed by a factor of 1.06). So, despite little changes in the static structure strong changes of the tail motion and bending elasticity seem to be a consequence of drug molecules penetrating the free space between the lipids.

We compared our bending rigidity values with other existing work and established a general trend in membrane rigidity changes induced by structurally similar drugs on phospholipid vesicles in the fluid phase. Since Acetaminophen toxicity has been related to multiple cellular functions such as oxidative stress, lipid peroxidation[62], etc., the changes in membrane dynamics and fluidity may be directly or indirectly connected to these cellular functions hence should not be ignored. This also opens up therapeutic avenues to explore the usage of clinically well-established "old" drugs such as Acetaminophen in cancer therapeutic drug delivery systems as an agent to manipulate membrane rigidity.[63]

## Supporting Information Description

Physicochemical details of Acetaminophen, Cryo-TEM additional images and analysis details, SAXS data from CAMD – lab X-ray source, NSE data modeling using ZG model, Impact of translational diffusion on NSE data, ZG decay rate $\Gamma_Q$ variation with and without the translational diffusion of the vesicles.

## Acknowledgment

The neutron scattering work is supported by the U.S. Department of Energy (DOE) under EPSCoR Grant No. DE-SC0012432 with additional support from the Louisiana Board of Regents. This work made use of the BioCryo facility of Northwestern University's NUANCE Center, which has received support from the SHyNE Resource (NSF ECCS-2025633), the IIN, and Northwestern's MRSEC program (NSF DMR-1720139). Access to the neutron spin echo spectrometer and small-angle scattering instruments was provided by the Center for High-Resolution Neutron Scattering, a partnership between the National Institute of Standards and Technology and the National Science Foundation under Agreement No. DMR-1508249. We acknowledge support from J. Krzywon for experiments at NG-7 SANS beamline. Research conducted at the Spallation Neutron Source (SNS) at Oak Ridge National Laboratory (ORNL) was sponsored by the Scientific User Facilities Division, Office of Basic Energy Sciences, U.S. DOE. The authors acknowledge Thomas Weiss from BL 4-2 at Stanford Synchrotron Radiation Lightsource for assisting with SAXS experiments (SSRL Proposal No. 5195). Use of the Stanford Synchrotron Radiation Lightsource, SLAC National Accelerator Laboratory, is supported by the U.S. Department of Energy, Office of Science, Office of Basic Energy Sciences under Contract No. DE-AC02-76SF00515. The SSRL Structural Molecular Biology Program is supported by the DOE Office of Biological and Environmental Research, and by the National Institutes of Health, National Institute of General Medical Sciences (including P41GM103393). The contents of this publication are solely the responsibility of the authors and do not necessarily represent the official views of NIGMS or NIH. SAXS data used in this publication were collected at the Small Angle X-ray Scattering (SAXS) beamline at the Center for Advanced Microstructures and Devices (CAMD). The authors thank Jiabo He at Tulane University for assisting with preliminary cryo-TEM experiments S.H. acknowledges financial support by the REU NSF CHE1660009.

## Disclaimer

## Table of Content Figure

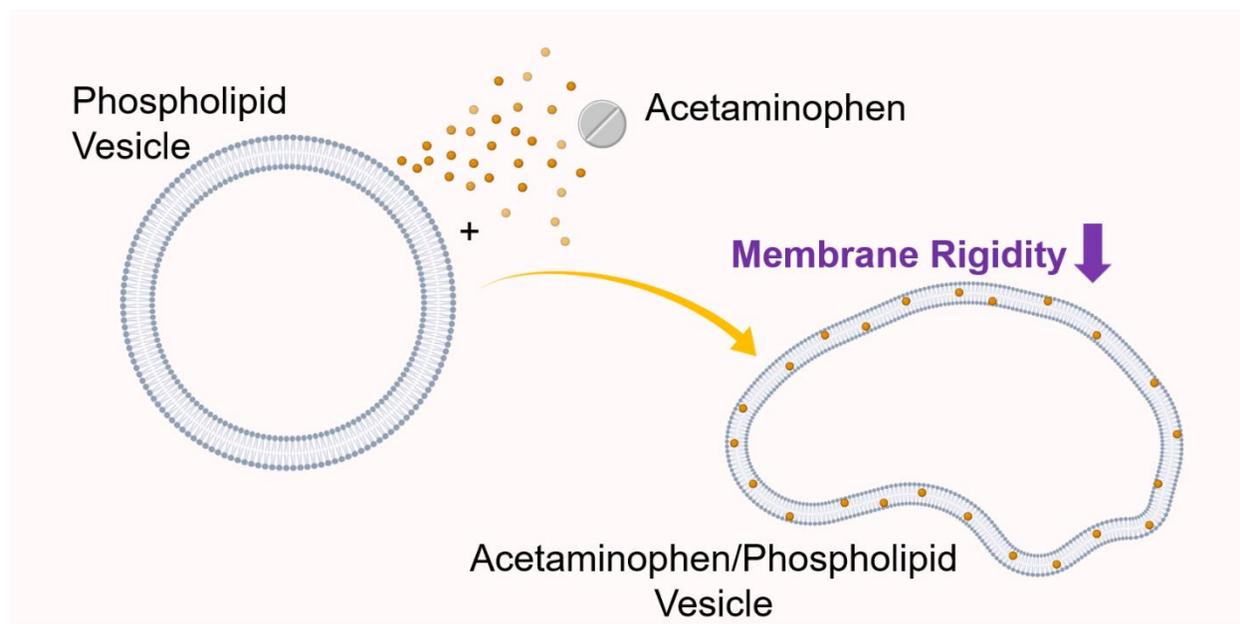